\documentclass[fleqn,usenatbib,useAMS]{mnras}

\usepackage{graphicx}	
\usepackage{amsmath}	
\usepackage{amssymb}	
\usepackage{multicol}        
\usepackage{bm}		
\usepackage{pdflscape}	
\usepackage{epsfig}

\usepackage[T1]{fontenc}
\usepackage{ae,aecompl}

\usepackage{newtxtext,newtxmath}
\hypersetup{pdfauthor={J. W. Broderick et al.}, pdftitle={Strong low-frequency radio flaring from Cygnus X-3 observed with LOFAR}, pdfkeywords={stars: individual: Cygnus X-3 -- radio continuum: stars -- ISM: jets and outflows -- X-rays: binaries}, bookmarksnumbered=true}

\title[LOFAR observations of Cygnus X-3 in outburst]{Strong low-frequency radio flaring from Cygnus X-3 observed with LOFAR}

\author[J. W. Broderick et al.]{J. W. Broderick,$^{1}$\thanks{E-mail: jess.broderick@curtin.edu.au}
T. D. Russell,$^{2,3}$
R. P. Fender,$^{4}$
S. A. Trushkin,$^{5,6}$ 
D. A. Green,$^{7}$ \newauthor
J. Chauhan,$^{1}$ 
N. A. Nizhelskij,$^{5}$ 
P. G. Tsybulev,$^{5}$ 
N. N. Bursov,$^{5}$ 
A. V. Shevchenko,$^{5}$ \newauthor
G. G. Pooley,$^{7}$\thanks{Deceased.}
D. R. A. Williams,$^{4,8}$
J. S. Bright,$^{4}$  
A. Rowlinson$^{2,9}$ and
S. Corbel$^{10,11}$
\\
$^{1}$International Centre for Radio Astronomy Research, Curtin University, GPO Box U1987, Perth, WA 6845, Australia\\
$^{2}$Anton Pannekoek Institute for Astronomy, University of Amsterdam, Science Park 904, NL-1098 XH Amsterdam, The Netherlands\\
$^{3}$INAF, Istituto di Astrofisica Spaziale e Fisica Cosmica, Via U. La Malfa 153, I-90146 Palermo, Italy\\
$^{4}$Astrophysics, Department of Physics, University of Oxford, Keble Road, Oxford OX1 3RH, UK\\
$^{5}$Special Astrophysical Observatory RAS, Nizhnij Arkhyz, Karachay-Cherkessia 369167, Russian Federation\\
$^{6}$Kazan Federal University, Kazan 420008, Russian Federation\\ 
$^{7}$Astrophysics Group, Cavendish Laboratory, 19 J. J. Thomson Avenue, Cambridge CB3 0HE, UK\\
$^{8}$Jodrell Bank Centre for Astrophysics, School of Physics and Astronomy, The University of Manchester, Manchester M13 9PL, UK\\
$^{9}$ASTRON, the Netherlands Institute for Radio Astronomy, Oude Hoogeveensedijk 4, NL-7991 PD Dwingeloo, The Netherlands\\
$^{10}$AIM, CEA, Universit{\'e} de Paris, Universit{\'e} Paris-Saclay, CNRS, F-91191 Gif-sur-Yvette, France\\
$^{11}$Station de Radioastronomie de Nan\c{c}ay, Observatoire de Paris, PSL Research University, CNRS, Univ. Orl\'{e}ans, F-18330 Nan\c{c}ay, France\\ 
}
\date{Accepted XXX. Received YYY; in original form ZZZ}

\pubyear{2021}

\begin{document}
\label{firstpage}
\pagerange{\pageref{firstpage}--\pageref{lastpage}}
\maketitle

\begin{abstract}
We present Low-Frequency Array (LOFAR) $143.5$-MHz radio observations of flaring activity during 2019 May from the X-ray binary Cygnus~X-3. Similar to radio observations of previous outbursts from Cygnus~X-3, we find that this source was significantly variable at low frequencies, reaching a maximum flux density of about $5.8$ Jy. We compare our LOFAR light curve with contemporaneous observations taken at $1.25$ and $2.3$ GHz with the RATAN-600 telescope, and at $15$ GHz with the Arcminute Microkelvin Imager (AMI) Large Array. The initial $143.5$-MHz flux density level, $\sim$$2$ Jy, is suggested to be the delayed and possibly blended emission from at least some of the flaring activity that had been detected at higher frequencies before our LOFAR observations had begun. There is also evidence of a delay of more than four days between a bright flare that initially peaked on May 6 at $2.3$ and $15$ GHz, and the corresponding peak ($\gtrsim 5.8$ Jy) at $143.5$ MHz. From the multi-frequency light curves, we estimate the minimum energy and magnetic field required to produce this flare to be roughly $10^{44}$ erg and $40$ mG, respectively, corresponding to a minimum mean power of $\sim$$10^{38}$~erg~s$^{-1}$. Additionally, we show that the broadband radio spectrum evolved over the course of our observing campaign; in particular, the two-point spectral index between $143.5$ MHz and $1.25$ GHz transitioned from being optically thick to optically thin as the flare simultaneously brightened at $143.5$ MHz and faded at GHz frequencies. 
\end{abstract}

\begin{keywords}
stars: individual: Cygnus X-3 -- radio continuum: stars -- ISM: jets and outflows -- X-rays: binaries
\newline\newline  
\end{keywords}



\section{Introduction}\label{introduction}

Cygnus~X-3 \citep[][]{giacconi67} is a Galactic high-mass X-ray binary, consisting of a Wolf--Rayet star \citep*[][]{vankerkwijk92,vankerkwijk96,fender99,kochmiramond02} in orbit with a black hole or neutron star primary compact object. The nature of the primary has been the subject of significant analysis \citep*[e.g.][]{hanson00,stark03,hjalmarsdotter08,szostek08,vilhu09,shrader10,zdziarski13,koljonen17}. Cygnus~X-3 is located at Galactic coordinates ($l$, $b$) = ($79$\fdg$8$, $+0$\fdg$7$), and the distance to the source has been estimated to be $\approx$~$3.4$--$10$ kpc, with a preferred distance $\approx$~$7$ kpc \citep*[e.g.][]{predehl00,ling09,mccollough16}.

At radio wavelengths, Cygnus~X-3 was first detected by \citet[][]{braes72}, who also reported significant variability in their $1.4$-GHz observations. Radio flaring from Cygnus~X-3 has now been studied for nearly half a century. At frequencies above $1$ GHz, the quiescent flux density level is $\sim$$100$ mJy \citep[e.g.][]{waltman94,zdziarski18}. However, during giant radio outbursts at these frequencies, flux densities of up to $\sim$$20$ Jy have been observed \citep*[e.g.][]{gregory72a,gregory72b,johnston86,waltman95,fender97,tsuboi08,trushkin08a,corbel12,trushkin17a,trushkin17b,egron17,egron21}. During such outbursts, high-resolution radio observations, particularly  very-long-baseline-interferometric measurements, have either strongly suggested the presence of relativistic jets in this system, or clearly resolved these structures \citep*[e.g.][]{geldzahler83,spencer86,molnar88,schalinski95,mioduszewski01,marti01,millerjones04,tudose07}. Immediately prior to the periods of bright radio flaring, Cygnus~X-3 enters a quenched, `hypersoft' state \citep[e.g.][]{mccollough99,szostek08,koljonen10,koljonen18}, where the flux density at GHz frequencies is typically from about one mJy to a few tens of mJy \citep[e.g.][]{hjellming72,waltman94,waltman96,fender97}. 

Despite there being a number of well-studied radio outbursts from Cygnus~X-3, not many low-frequency ($< 400$ MHz) radio observations have been taken during these events. Following the first recorded giant outburst at GHz frequencies in 1972 September (\citealt[][]{gregory72b} et seq.), \citet[][]{bash73} conducted $365$-MHz monitoring observations with the University of Texas Broadband Synthesis Interferometer. More recently, low-frequency observing campaigns, which also included simultaneous or contemporaneous higher-frequency observations, took place in 2001 September, 2006 May--June, 2007 June, and 2008 April and December  \citep*{millerjones04,millerjones07,pal07,pal08,pal09,patra15}. These low-frequency observations were conducted at $74$ and $330$ MHz with the Karl G. Jansky Very Large Array (VLA), at $140$ MHz with the Westerbork Synthesis Radio Telescope (WSRT), and at $244$ and $325$ MHz with the Giant Metrewave Radio Telescope (GMRT). Cygnus~X-3 is also a variable source at low frequencies, with measured flux densities up to $7.9$ Jy (at $365$ MHz; \citealt[][]{bash73}), although only $74$-MHz upper limits have been reported \citep[][]{millerjones04}. 

During the 1972, 2001, 2006, 2007 and 2008 flaring events, the turnover frequency of the radio spectrum was seen to shift to lower values as each outburst progressed (\citealt*[][]{marsh74} and references therein; \citealt{millerjones04,millerjones07,pal08,pal09,trushkin08b,patra15}). The radio spectral index, $\alpha$\footnote{\label{footnote:spectral index}$S_{\nu} \propto \nu^{\alpha}$, where $S_{\nu}$ is the flux density at frequency $v$. Moreover, in this paper, we use the notation $\alpha^{\nu_2}_{\nu_1}$ to denote a two-point spectral index between $\nu_1$ and $\nu_2$ MHz.}, subsequently evolved as well. For example, over a period of nine days during the 2006 outburst, the two-point spectral index $\alpha^{614}_{244}$ changed in value from $1.89$ to $-0.98$, i.e. from being very inverted and optically thick, to steep and optically thin \citep[][]{pal09}. The radio spectral evolution of this outburst, as well as the 1972, 2001, 2007 and 2008 events, was discussed in terms of an often-invoked synchrotron bubble model where the emitting plasmons become progressively less optically thick as they propagate outwards and expand \citep[e.g.][]{vanderlaan66,hjellming88,ball93}. Mechanisms were also investigated to account for the spectral turnover, namely synchrotron self-absorption, free--free absorption, the Razin--Tsytovich effect, and a low-frequency cutoff in the synchrotron spectrum. 

Another study of interest was carried out by \citet[][]{pandey07}, where ten simultaneous $235$- and $610$-MHz GMRT observations of Cygnus~X-3 were taken between 2003 July and 2005 January. Although no bright flares were detected, the source was found to be variable at $235$ MHz over the time-scale of the observing campaign: the $235$-MHz flux density varied between $4.9$ and $49$ mJy, with a mean of $18$ mJy. Inverted radio spectra were reported for all observations, with the two-point spectral index $\alpha^{610}_{235}$ per epoch ranging from $0.09$ to $1.23$. 

Observations of outbursting X-ray binaries have been carried out with several facilities from the current generation of low-frequency radio telescopes: the Low-Frequency Array (LOFAR), the Murchison Widefield Array (MWA), and the VLA Low-Band Ionosphere and Transient Experiment (VLITE). In combination with higher-frequency radio observations, these data have allowed valuable constraints to be placed on the broadband spectral properties of the flares. Hence, such observations have 
offered new insights into the physical processes responsible for the observed emission, as well as the mechanisms responsible for any spectral turnover (e.g. \citealt[][]{kassim15,broderick15,broderick18,polisensky18,chauhan19}; also see \citealt{marcote16} for the case of low-frequency variability from a gamma-ray binary). 

In 2019, an ideal opportunity arose to obtain new low-frequency observations of a giant outburst from Cygnus~X-3. Monitoring at five separate frequencies between $4.6$ and $30$ GHz with the RATAN-600 telescope indicated that the radio emission from Cygnus~X-3 had become quenched on 2019 February 17 \citep[][]{trushkin19a}. Following the quenched phase, radio flaring was observed from 2019 April 17 (observations between $1.25$ and $37$ GHz with RATAN-600, the Nasu Telescope Array, the Arcminute Microkelvin Imager (AMI) Large Array, and the Mets\"{a}hovi Radio Observatory), with activity continuing into 2019 May (\citealt[][]{koljonen19,tsubono19a,trushkin19b}; also see \citealt[][]{piano19a,piano19b} for detections of gamma-ray flaring, and \citealt{choudhury19} for X-ray monitoring). 

In this paper, we present LOFAR monitoring observations of Cygnus~X-3 during the 2019 April--May event. In Section~\ref{observations}, we describe our LOFAR observations and calibration method. Section~\ref{flux scale accuracy} discusses imaging, and in Section~\ref{light curve} a LOFAR light curve for Cygnus~X-3 is presented, as well as the associated variability statistics. In Section~\ref{discussion} we compare the LOFAR data with contemporaneous radio observations taken at higher frequencies, discussing plausible scenarios for the observed data. We then report our conclusions in Section~\ref{conclusions}. All uncertainties in this paper are quoted at the $68$ per cent confidence level. 

\section{LOFAR observations and data reduction}\label{observations}

\begin{table*}
\begin{minipage}{0.8\textwidth}
 \centering
  \caption{LOFAR observing log, $143.5$-MHz flux densities for Cygnus~X-3, root-mean-square (rms) noise levels in the vicinity of Cygnus~X-3, multiplicative factors used to correct the flux densities and rms noise levels (Section~\ref{flux scale accuracy}), and details of the synthesized beams. The modified Julian date (MJD) for each run corresponds to the midpoint of the $48$-min observation. The quoted internal uncertainties for the flux densities are at the $1\sigma$ level, and do not include systematic effects associated with the accuracy of the TGSS flux density scale; see Section~\ref{flux scale accuracy} and Table~\ref{table:alphas} for further details. PA is the position angle of the synthesized beam, measured north through east.}
  \begin{tabular}{ccccccc}
    \hline
Run & Date & MJD & $S_{143.5}$ & Noise level & Applied flux density & Angular resolution and PA \\
& & & (Jy) & (mJy beam$^{-1}$) & correction factor ($\times$) & (arcsec$^{2}$; \degr) \\
\hline
1 & 2019 May 2 & 58605.344  & $2.08 \pm 0.37$ & $51$ & $2.33 \pm 0.37$ & $111 \times	88$; $80$
  \\
2 & 2019 May 3 & 58606.248 & $1.68 \pm 0.23$ & $62$ & $1.76 \pm 0.20$ & $98 \times 92$; $-54$  \\
3 & 2019 May 4 & 58607.318 & $2.50 \pm 0.58$ & $64$ & $2.27 \pm 0.51$ & $106 \times	90$; $89$  \\
4 & 2019 May 5 & 58608.225 & $1.84 \pm 0.21$ & $58$ & $1.44 \pm 0.13$ & $98 \times 92$; $-45$  \\
5 & 2019 May 6 & 58609.246 & $2.48 \pm 0.31$ & $88$ & $1.84 \pm 0.19$ & $98 \times 92$; $-58$  \\
6 & 2019 May 7 & 58610.246 & $2.62 \pm 0.63$ & $128$ & $2.28 \pm 0.50$ & $111 \times 99$; $-47$ \\
7 & 2019 May 10 & 58613.246 & $5.24 \pm 0.78$ & $154$ & $2.03 \pm 0.27$ & $99 \times 91$; $-63$  \\
8 & 2019 May 16 & 58619.246 & $5.82 \pm 0.72$ & $131$ & $1.80 \pm 0.20$ & $100 \times	91$; $-72$  \\
\hline
\end{tabular}
\label{table:obs}
\end{minipage}
\end{table*}

The LOFAR telescope is described in \citet[][]{vanhaarlem13}. Our observing campaign (approved target-of-opportunity observations under project code LC11\_021; total observing time of $8$ h) commenced approximately $15$ days after the initial detection of flaring in the radio band. Eight observations were made in total, with an approximately daily cadence for the first six runs, followed by gaps of three and six days for the final two runs.  Observation dates are provided in Table~\ref{table:obs}. 

The observations were taken using the high-band antennas (HBA) in the `HBA Dual Inner' configuration. Each observation used $38$ LOFAR stations: $24$ core stations (each with two sub-stations or `ears') and $14$ remote stations. All observations consisted of $48$ min on target, preceded or followed by a $10$-min scan of a primary calibrator, which was either 3C~295 or 3C~48. Given that Cygnus~X-3 can reach flux densities of a Jy or more at low frequencies during an outburst (Section~\ref{introduction}), the duration of the target scans was a conservative choice related to expected challenges in imaging each run (as discussed below in this section).    

The data were recorded across the frequency range $115$--$189$ MHz, with each of the $380$ sub-bands having a bandwidth of $195.3$ kHz. However, in this study, we made use of the $240$ sub-bands between $120$ and $167$ MHz (central frequency $143.5$ MHz) to reduce issues with radio-frequency interference (RFI), and optimize the bandpass sensitivity. At $143.5$ MHz, the primary beam full width at half maximum (FWHM) was $4$\fdg$0$ \citep[][]{vanhaarlem13}. 

The native temporal and frequency resolutions were $1$~s and $64$ channels per sub-band, respectively. After preprocessing, including the removal of RFI with {\sc aoflagger} \citep*[][]{offringa10,offringa12a,offringa12b}, as well as {\sc dysco} compression \citep[][]{offringa16}, our data had temporal and frequency resolutions of $4$~s and $4$ channels per sub-band, respectively. We then calibrated the data using {\sc prefactor} \citep[][]{vanweeren16,williams16,degasperin19}\footnote{\url{https://github.com/lofar-astron/prefactor}}. We used calibrator source models from \citet[][]{scaife12}, while the target field sky model used for phase-only calibration was based on data from the $147.5$-MHz TIFR GMRT Sky Survey First Alternative Data Release \citep[TGSS ADR1; angular resolution $25$ arcsec;][]{intema17}. Moreover, the TGSS model included all sources with flux density $S_{147.5} > 300$ mJy within $5$\degr\,of Cygnus~X-3.  

Imaging Cygnus~X-3 with LOFAR is challenging due to (i) the field containing a variety of complex extended emission \citep[e.g.][]{millerjones07}, and (ii) the very bright source, Cygnus~A, is only $6\fdg2$ away \citep[$150$-MHz flux density $10.7$ kJy;][]{mckean16}. During preprocessing, an attempt to use the `demixing' algorithm \citep*[][]{vandertol07} with standard settings to remove the response from Cygnus~A resulted in final processed images that contained no sources. A full direction-dependent calibration would have best ameliorated the problem, and would also have been an excellent test case for LOFAR fields in close proximity to very bright sources. However, this had the potential to become a very time-consuming process. Because we had the additional important consideration of rapidly reporting initial results and ongoing observations to the community \citep[][]{broderick19}, we chose instead to process and analyse data sets where Cygnus A had not been removed from the visibilities, and it is these results that we present in this paper. We therefore added a model of Cygnus A from the {\sc prefactor} software, comprising $12$ Gaussian or point-like components, to our target field sky model. Not fully solving for Cygnus A resulted in significant sensitivity and angular resolution penalties, and extra care was also needed to quantify the accuracy of the flux density scale (Section~\ref{flux scale accuracy} and Appendix~\ref{appendix1}). Nonetheless, we will demonstrate in Section~\ref{light curve} that we were still able to reliably determine the flux densities of Cygnus~X-3 and detect statistically significant variability. 

\section{Imaging}\label{flux scale accuracy}

\begin{figure}
\epsfig{file=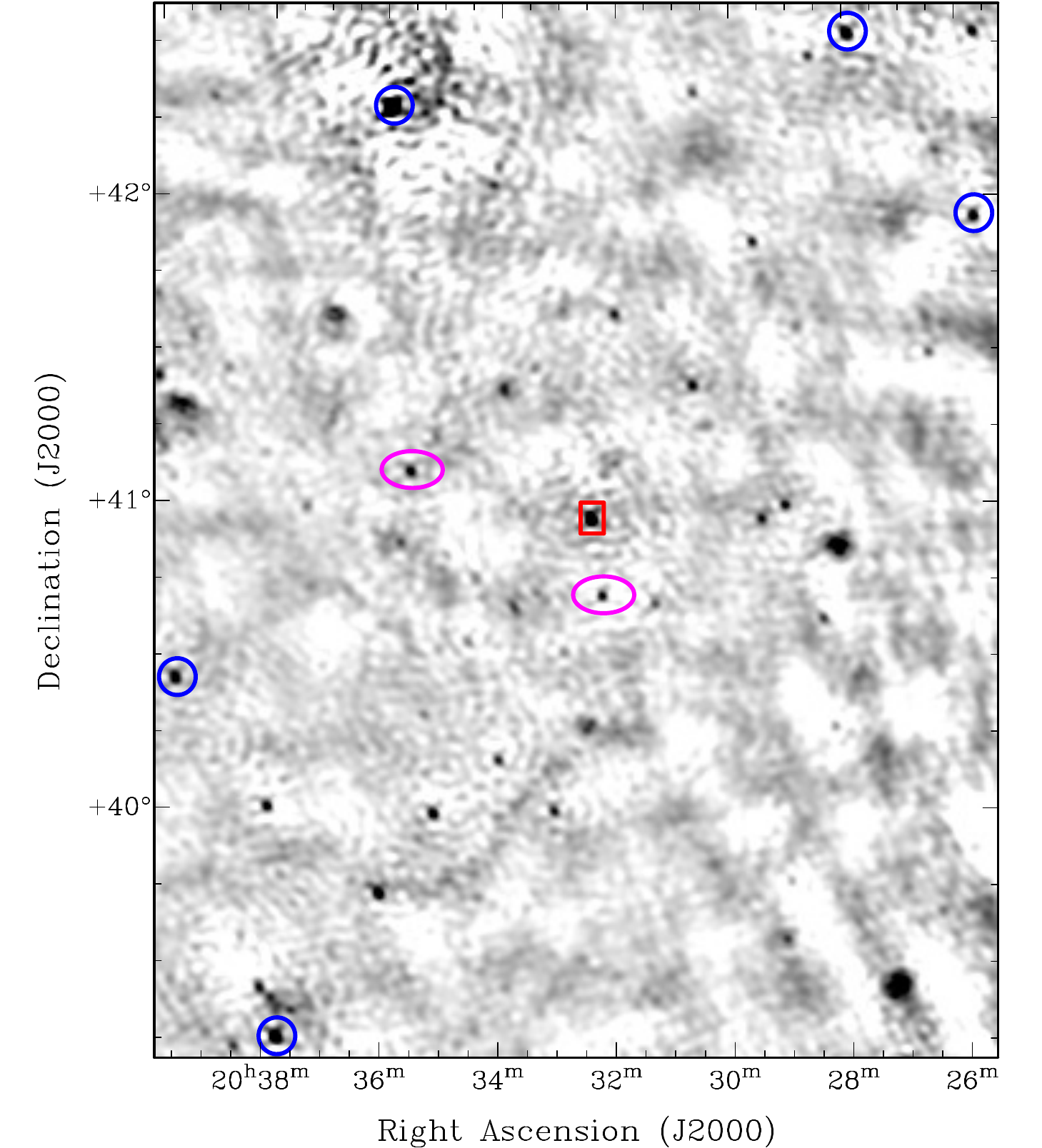,width=8.25cm}
\caption{The inner part of our $143.5$-MHz map from the first observation on 2019 May 2. Cygnus~X-3 is marked with the rectangle, sources used to bootstrap the flux density scale from TGSS (Appendix~\ref{appendix1}) are indicated with circles, and the ellipses mark the two sources whose light curves we show in Figure~\ref{fig:appendix}. Cygnus~A is $6$\fdg$2$ to the west of the centre of the map (not visible in this zoomed-in view). This image has a synthesized beam of $111 \times 88$ arcsec$^{2}$ (PA $80$\degr), and an rms noise level of $51$ mJy beam$^{-1}$ near Cygnus~X-3. There is an offset between our astrometry and that of TGSS, such that the LOFAR positions are approximately $30$ arcsec to the west-north-west, on average. However, as this offset, as well as similar offsets in the images for the other runs, did not affect any of the analysis presented in this study, no astrometric corrections were applied to the LOFAR data.}
\label{fig:example}
\end{figure}

We used {\sc wsclean} \citep[][]{offringa14}\footnote{\url{https://sourceforge.net/projects/wsclean}} to image our data; an example image is shown in Figure~\ref{fig:example}. A projected baseline range of $100$--$1500$$\lambda$ was implemented; the lower cutoff reduced the contribution from large-scale, diffuse Galactic emission in the field, while the upper cutoff ensured that the angular resolution remained relatively coarse, helping to reduce dynamic range issues associated with Cygnus~A. We also used multi-frequency deconvolution, grouping the data into six channels of $40$ sub-bands each (bandwidth $7.8$ MHz per channel). A Briggs robust weighting parameter of $-0.5$ was selected for all runs except Run 6. In the case of Run 6, we found that we needed to use a lower robust parameter ($-1.25$) to achieve an angular resolution similar to the other runs. This was likely related to a difference in the weights of the visibilities in the measurement sets for this run, prior to any imaging taking place. It was not entirely clear why the weights were different in this case, but we believe that an appropriate flux density scale correction (as discussed below in this section and in Appendix~\ref{appendix1}) allowed us to measure a reliable flux density for Run 6. 

The angular resolutions for each of the runs are given in Table~\ref{table:obs}. The average resolution was about $100$ arcsec. Flux density changes resulting from the differing angular resolutions in each of the runs were insignificant compared to the fitting and calibration uncertainties. Henceforth, all results presented are at the native angular resolution of each data set.

Generally speaking, LOFAR flux density calibration requires a bootstrapping step to correct the flux density scale \citep[e.g.][]{hardcastle16}. For our data, we used TGSS for bootstrapping. The procedure is described in Appendix~\ref{appendix1}, and the correction factors are given in Table~\ref{table:obs}. After this step, we then used {\sc pybdsf} \citep[][]{mohan15} to measure the flux density of Cygnus~X-3 in each LOFAR map. The corrected flux densities are given in Table~\ref{table:obs}. Each flux density uncertainty was determined by combining, using appropriate error propagation, the uncertainty in the correction factor and the flux density fitting uncertainty from {\sc pybdsf}. As we discuss in Appendix~\ref{appendix2}, despite the relatively coarse angular resolution, possible blending from nearby extended emission was negligible. 

The corrected rms noise levels in the vicinity of Cygnus~X-3 ranged from $51$ to $154$ mJy beam$^{-1}$ (Table~\ref{table:obs}). Such noise levels are, at best, over an order of magnitude higher than the expected classical confusion limits at the angular resolutions and central frequency of our observations (e.g. \citealt{franzen16,franzen19}, and references therein). Given the challenge of calibrating and imaging our observations, we did not attempt to self-calibrate the data using source models derived from the LOFAR maps.

A potential advantage of our wide-bandwidth, low-frequency radio data was to obtain information on the radio spectrum of Cygnus X-3 across the LOFAR high band. However, inspection of the $7.8$-MHz bandwidth images produced by {\sc wsclean} as part of the multi-frequency deconvolution process for each run indicated that there were systematic effects across the bandpass, such that all sources in the field of view appeared to have inverted in-band spectra. While we effectively corrected for this on average with our bootstrapping procedure, further corrections across the band were hampered by a paucity of nearby, sufficiently bright sources with literature flux densities at frequencies both below and above the LOFAR high band. 

We conservatively assumed that the TGSS flux density scale accuracy is $20$ per cent for our target field; see the discussion in Appendix~\ref{appendix1}. In Section~\ref{light curve}, we analyse the Cygnus X-3 LOFAR flux densities, which were all bootstrapped to the TGSS flux density scale in the same way, and therefore an additional systematic calibration uncertainty will not affect the main conclusions. In Section~\ref{discussion}, however, we include the uncertainty in the low-frequency absolute flux density scale in our multi-frequency analysis.

\hspace{1cm}

\section{Cygnus X-3 light curve}\label{light curve}

\begin{figure*}
\begin{minipage}{1.0\textwidth}
\centering
\includegraphics[height=7.5cm]{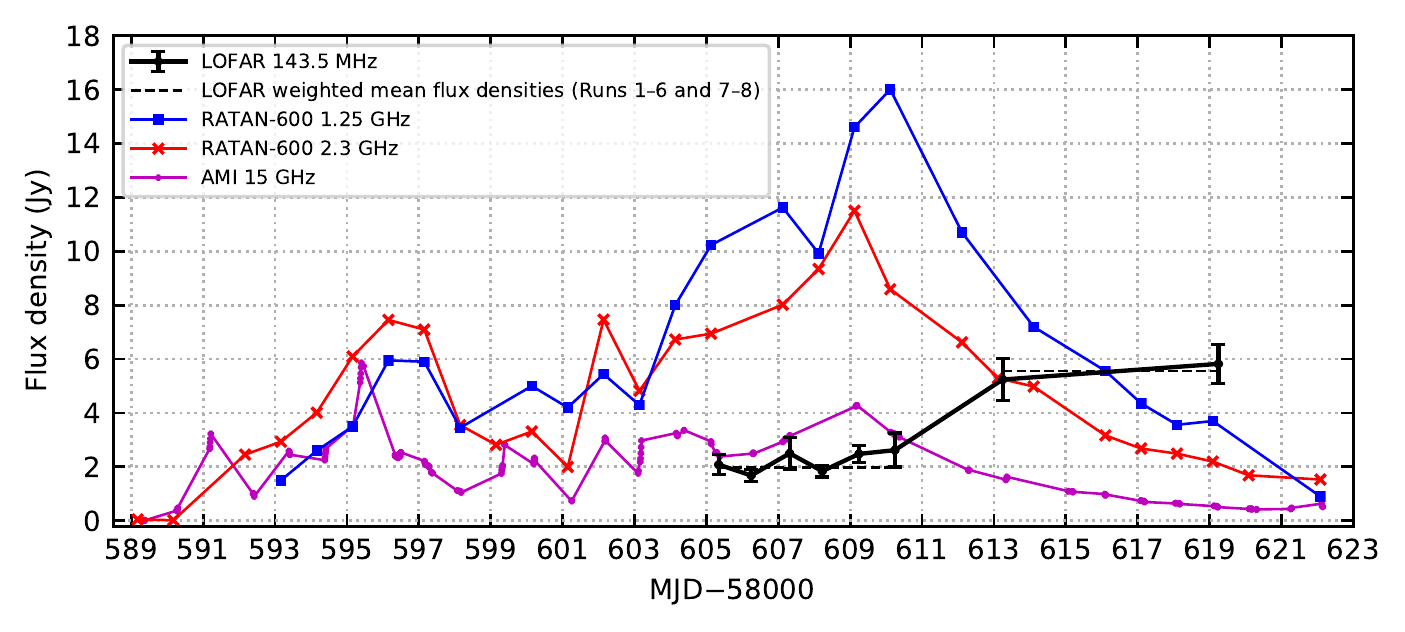}
\end{minipage}%
\caption{Our LOFAR light curve for Cygnus~X-3, using the data from Table~\ref{table:obs}. The two dashed lines are the inverse-variance-weighted mean flux densities for Runs 1--6 and Runs 7--8, respectively. We have also plotted $1.25$-GHz, $2.3$-GHz and $15$-GHz light curves from the contemporaneous RATAN-600 and AMI monitoring programmes (Section~\ref{light curve comparison}). The error bars for the LOFAR data points are $\pm1\sigma$. For the purposes of clarity, we have not plotted the uncertainties for the RATAN-600 and AMI flux densities; the uncertainties are $100$--$300$ mJy at $1.25$ GHz, and $5$ per cent at both $2.3$ and $15$ GHz. The $2.3$-GHz data point for MJD $-$ 58000 $=$ 590.171 is an upper limit ($<20$ mJy). The assumed $20$ per cent uncertainty in the absolute flux density scale of both TGSS and our bootstrapped LOFAR data (Section~\ref{flux scale accuracy} and Appendix~\ref{appendix1}), not accounted for in the plotted error bars, could result in a systematic shift of the LOFAR light curve.}
\label{fig:light curve}
\end{figure*}

Our $143.5$-MHz light curve for Cygnus~X-3 is shown in Figure~\ref{fig:light curve}. The source was bright and significantly detected in all eight observations, well above the quiescent flux density level at this frequency. Indeed, we inspected the $147.5$-MHz TGSS image products and calculated a $3\sigma$ upper limit for the quiescent flux density of about $30$ mJy beam$^{-1}$. Furthermore, the $235$- and $610$-MHz results from \citet[][]{pandey07} suggest that the average quiescent baseline at $143.5$ MHz is well below this TGSS upper limit (see Section~\ref{introduction}), as do recently published $325$- and $610$-MHz GMRT flux densities \citep[][]{benaglia20a,benaglia20b,benaglia21}. 

Within the uncertainties, Cygnus~X-3 had an approximately constant flux density from May 2--7 (MJD 58605--58610), but then brightened between May 7 and 10 (MJD 58610 and 58613). The inverse-variance-weighted mean flux densities of Runs 1--6 and 7--8 were $1.98 \pm 0.12$ and $5.55 \pm 0.53$ Jy, respectively, where each uncertainty is the standard error of the weighted mean (Figure~\ref{fig:light curve}). Therefore, the average flux density increased by a factor of $2.8 \pm 0.3$ between these two observational subsets. The latter mean flux density was also over a factor of two brighter than the peak $140$-MHz flux density of $2.3$ Jy during the 2006 May outburst (\citealt[][]{millerjones07}; also see Section~\ref{introduction}).  

Following, for example, \citet*[][]{kesteven76}, \citet[][]{gaensler00} and \citet[][]{bell14}, the $\chi^2$ probability that Cygnus~X-3 did not vary over the course of our observing campaign was calculated using the statistic 
\begin{equation}\label{eqnvar1}
    x^2 = \sum\limits^{8}_{i=1}\frac{(S_i - \widetilde{S})^{2}}{\sigma^{2}_i},
\end{equation}
where $S_i$ is the LOFAR flux density from the $i$th observation (Table~\ref{table:obs}), $\sigma_i$ the uncertainty in the flux density from the $i$th observation, and $\widetilde{S}$ the inverse-variance-weighted mean of all eight runs ($2.17$ Jy). From Equation~\ref{eqnvar1}, $x^2 = 50.1$. Assuming normally distributed uncertainties, $x^2$ follows a $\chi^2$ distribution with $8-1=7$ degrees of freedom in this case. The probability that $\chi^2 > 50.1$ for $7$ degrees of freedom is $1.4 \times 10^{-8}$. This is a highly significant $P$-value, providing strong statistical evidence that Cygnus~X-3 varied over the length of our observing campaign. 

The debiased modulation index, $m_{\rm d}$, was also used to quantify the relative variability of Cygnus~X-3 (e.g. \citealt[][]{bell14} and references therein). In the case of our data set, it was calculated using the expression
\begin{equation}\label{eqnvar3}
 m_{\rm d} = \frac{1}{\overline{S}} \sqrt{\frac{\sum_{i=1}^{8}(S_{i} - \overline{S})^{2} - \sum_{i=1}^{8}\sigma_{i}^{2}}{8}}, 
\end{equation}
where the unweighted mean flux density $\overline{S} = 3.03$ Jy. From Equation~\ref{eqnvar3}, $m_{\rm d} =$ 46 per cent.  

\section{Properties of the 2019 May bright flare}\label{discussion}

\subsection{Comparison with contemporaneous, higher-frequency data}\label{light curve comparison} 

\begin{table*}
\begin{minipage}{1.00\textwidth}
 \centering
  \caption{Flux densities, two-point spectral indices and fitted spectral indices for each LOFAR run. As we were directly comparing the LOFAR flux densities with higher-frequency data points, we added the $20$ per cent flux density scale uncertainty at $143.5$ MHz in quadrature with the calibration and fitting uncertainties described in Section~\ref{flux scale accuracy} and reported in Table~\ref{table:obs}. The flux densities at GHz frequencies were linearly interpolated to the MJDs of our LOFAR observations. Further details can be found in Section~\ref{light curve comparison2}.}
  \begin{tabular}{ccrrrrrrrr}
    \hline
\multicolumn{1}{c}{Run} & \multicolumn{1}{c}{MJD} & \multicolumn{1}{c}{$S_{143.5}$} & \multicolumn{1}{c}{$S_{1250}$} & \multicolumn{1}{c}{$S_{2300}$} & \multicolumn{1}{c}{$S_{15000}$} &  \multicolumn{1}{c}{$\alpha^{1250}_{143.5}$} & \multicolumn{1}{c}{$\alpha^{2300}_{1250}$} & \multicolumn{1}{c}{$\alpha^{15000}_{2300}$} & \multicolumn{1}{c}{$\alpha_{\rm fitted}$ and $\chi^2_{\rm red}$} \\  
& & \multicolumn{1}{c}{(Jy)} & \multicolumn{1}{c}{(Jy)} & \multicolumn{1}{c}{(Jy)} & \multicolumn{1}{c}{(Jy)} & & & & \multicolumn{1}{c}{($1250$--$15000$ MHz)} \\
  \hline
1 & 58605.344  & $2.1 \pm 0.6$ & $10.4 \pm 0.2$ & $7.1 \pm 0.4$ & $2.45 \pm 0.12$ & $0.74 \pm 0.13$ & $-0.63 \pm 0.10$ & $-0.57 \pm 0.04$ & $-0.58 \pm 0.02$; $0.21$ \\
2 & 58606.248 & $1.7 \pm 0.4$ & $11.0 \pm 0.2$  & $7.5 \pm 0.4$ & $2.49 \pm 0.12$  & $0.86 \pm 0.11$ & $-0.63 \pm 0.09$ & $-0.59 \pm 0.04$ & $-0.60 \pm 0.02$; $0.11$ \\
3 & 58607.318 & $2.5 \pm 0.8$ & $11.3 \pm 0.2$ & $8.3 \pm 0.4$ & $3.17 \pm 0.16$ & $0.70 \pm 0.15$ & $-0.51 \pm 0.08$ & $-0.51 \pm 0.04$ & $-0.51 \pm 0.02$; $0.0042$ \\
4 & 58608.225 & $1.8 \pm 0.4$ & $10.4 \pm 0.2$ & $9.6 \pm 0.5$ & $3.68 \pm 0.18$ & $0.81 \pm 0.10$ & $-0.13 \pm 0.09$ & $-0.51 \pm 0.04$ & $-0.41 \pm 0.02$; $8.7$ \\
5 & 58609.246 & $2.5 \pm 0.6$ & $14.8 \pm 0.3$ & $11.1 \pm 0.6$ & $4.22 \pm 0.21$ & $0.82 \pm 0.11$ & $-0.47 \pm 0.09$ & $-0.52 \pm 0.04$ & $-0.50 \pm 0.02$; $0.12$ \\
6 & 58610.246 & $2.6 \pm 0.8$ & $15.7 \pm 0.3$ & $8.5 \pm 0.4$ & $3.16 \pm 0.16$ & $0.83 \pm 0.14$ & $-1.01 \pm 0.08$ & $-0.53 \pm 0.04$ & $-0.66 \pm 0.02$; $22$ \\
7 & 58613.246 & $5.2 \pm 1.3$ & $8.7 \pm 0.3$ & $5.2 \pm 0.3$ & $1.55 \pm 0.08$ & $0.24 \pm 0.12$ & $-0.84 \pm 0.11$ & $-0.65 \pm 0.04$ & $-0.69 \pm 0.03$; $2.1$ \\
8 & 58619.246 & $5.8 \pm 1.4$ & $3.6 \pm 0.2$ & $2.1 \pm 0.1$ & $0.50 \pm 0.03$ & $-0.22 \pm 0.11$ & $-0.88 \pm 0.12$ & $-0.77 \pm 0.04$ & $-0.79 \pm 0.03$; $0.70$ \\
\hline
\end{tabular}
\label{table:alphas}
\end{minipage}
\end{table*}

We compared our LOFAR light curve with contemporaneous, independent observing campaigns carried out at higher frequencies. In particular, in Figure~\ref{fig:light curve}, we show the $1.25$- and $2.3$-GHz light curves from RATAN-600, as well as a $15$-GHz light curve from the AMI Large Array. Overviews of these facilities can be found in \citet[][]{khaikin72} (RATAN-600), as well as \citet[][]{zwart08} and \citet[][]{hickish18} (AMI). 

From the commencement of flaring, until a few days after the end of our LOFAR observations, there were $23$ RATAN-600 observations taken at $1.25$ GHz and $29$ at $2.3$ GHz. An initial description of the data can be found in \citet[][]{trushkin19b}. Mostly a daily observing cadence was used. The $1.25$-GHz observations had a beam size of approximately $12.9 \times 2.2$ arcmin$^2$, while the beam size at $2.3$ GHz was approximately $7.0 \times 1.2$ arcmin$^{2}$ (PA $0$\degr\:in both cases; see \citealt{majorova02} for further details of the RATAN-600 beam patterns). NGC~7027 was used as the primary flux density calibrator, and the flux density uncertainty per target measurement was $100$--$300$ mJy ($1.25$ GHz) or $5$ per cent ($2.3$ GHz). The full set of monitoring data, spanning seven separate frequencies from $1.25$ to $30$ GHz, will be presented in a future paper. 

A total of $212$ AMI monitoring scans are also shown in Figure~\ref{fig:light curve}. Typically, $5$ $\times$ $10$-min scans were taken per day, interleaved with short scans of the phase calibrator, J2052$+$3635. The observations had an angular resolution of approximately $30$ arcsec. Note that the telescope measures Stokes $I + Q$ (i.e. a single linear polarization). The primary calibrator was 3C~286, and additional small flux density corrections were made using J2052$+$3635. We have conservatively assumed an uncertainty of $5$ per cent for each Cygnus X-3 flux density measurement. The full set of AMI monitoring data will also be presented in a future paper. 

In Figure~\ref{fig:light curve}, we can see that there was bright and prolonged flaring from Cygnus X-3 at GHz frequencies during 2019 April--May, with multiple flare peaks at each frequency. As previously mentioned in Section~\ref{light curve}, the LOFAR flux density initially remained relatively stable, albeit at a heightened brightness compared to the normal quiescent level. This behaviour was very likely due to the associated delay expected between individual flares at high and low frequencies (see Section~\ref{introduction} and references therein, as well as further discussion in Section~\ref{predictions}). It is also evident that while the higher-frequency light curves peaked on MJDs 58609 and 58610 and then started to decay, the LOFAR light curve clearly exhibited a delayed flux density rise only after MJD 58610. We consider it unlikely that the $143.5$-MHz peak was at or very near the time of Run $7$ (MJD 58613), as the subsequent flare decay would have been almost flat initially, and markedly different to what was observed at GHz frequencies. However, due to the six-day gap between our final two observations (MJD 58613--58619), it is unclear whether the flare peaked between Runs 7 and 8, during Run 8, or after our last observation. 

\subsection{Broadband spectral characteristics}\label{light curve comparison2} 

We explored the spectral characteristics of the light curves (Table~\ref{table:alphas}). The RATAN-600 and AMI flux densities were linearly interpolated to the MJDs of the LOFAR data using the two nearest measurements in time. Relative to the closest measurement in time in each case, these adjustments were at most $21$ per cent, with a median of $2.0$ per cent and only two instances above $5.9$ per cent. Because our analysis would not be significantly improved by accounting for the uncertainties associated with these mostly small corrections, for simplicity we assumed that the interpolated RATAN-600 and AMI flux densities had the same absolute or relative uncertainties as described in Section~\ref{light curve comparison}. Furthermore, as we were directly comparing the LOFAR flux densities with GHz-frequency measurements, we added the aforementioned $20$ per cent absolute flux scale uncertainty from the TGSS bootstrapping process (Section~\ref{flux scale accuracy}) in quadrature with the previously reported uncertainty for each LOFAR flux density as given in Table~\ref{table:obs}. 

Table~\ref{table:alphas} includes the spectral indices determined from a single-power-law fit to the three data points between $1.25$ and $15$ GHz at the MJD of each LOFAR epoch. Consistent with the convention established in Footnote~\ref{footnote:spectral index}, we used the standard function 
\begin{equation}\label{eqnfit}
S_\nu = S_0 \left( \frac{\nu}{\nu_0} \right)^\alpha,
\end{equation}
where $S_{0}$ is the flux density at reference frequency $\nu_{0}$. We set $\nu_{0} = 8.125$ GHz, i.e. in the middle of our frequency range in linear space, but, for example, using the central frequency in log space, that is $4.33$ GHz, made no difference to the final results. The resulting fitted values of $\alpha$, along with the corresponding $\chi^2$ goodness-of-fit statistics (equivalent to the reduced $\chi^2$, $\chi^2_{\rm red}$, in this case, as there was only one degree of freedom per fit), are also given in Table~\ref{table:alphas}. Some of the $\chi^2_{\rm red}$ values suggest overestimated flux density uncertainties, although the formal standard deviation on this statistic is relatively large ($\sqrt{2}$) given our limited data points. Moreover, the probability that $\chi^2_{\rm red} \leq 0.0042$ (our smallest value, from Run 3) is still $5.2$ per cent. 

\begin{figure}
\epsfig{file=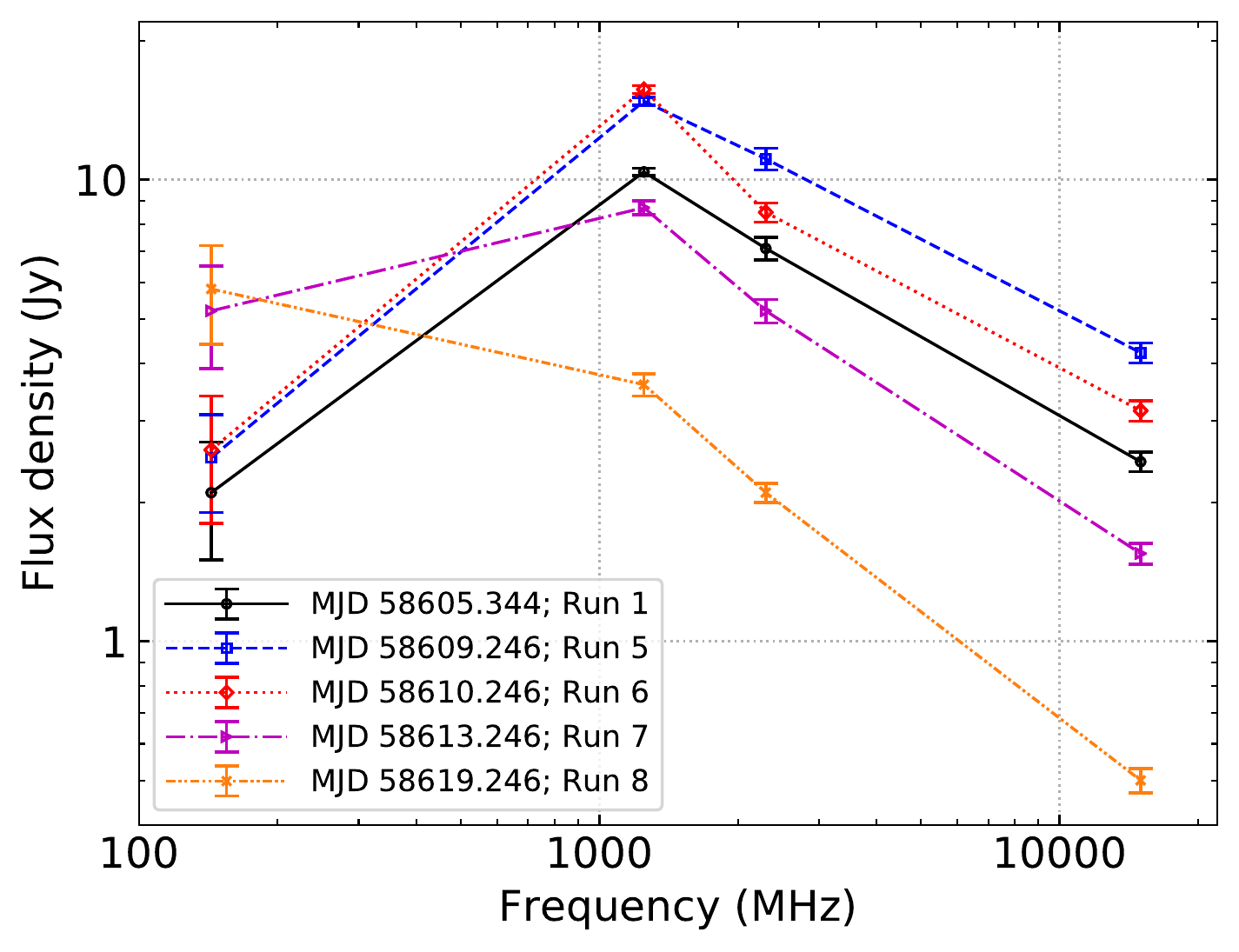,width=8.4cm}
\caption{Radio spectra for five selected runs: 1, 5, 6, 7 and 8. We have used the data presented in Table~\ref{table:alphas}; see the table caption for further details. Plotted lines are not fits to the data, but rather illustrate how the two-point spectral index changed as a function of frequency at a particular epoch. Run 1 shows the initial shape of the radio spectrum when our LOFAR observing campaign commenced, Runs 5 and 6 were very close to the peak of the flaring as measured at the higher frequencies, and Runs 7 and 8 are of particular interest because of the significant increase in the LOFAR flux density. The broadband radio spectrum clearly evolved over the course of the flare.}
\label{fig:seds}
\end{figure}

The two-point and fitted radio spectra in Table~\ref{table:alphas} varied significantly over the course of our observing campaign, both in terms of brightness and spectral slope/shape. We show this further in Figure~\ref{fig:seds}, where we have plotted broadband radio spectra at MJDs corresponding to Runs 1, 5, 6, 7 and 8. Initially, above $1.25$ GHz, the spectrum was well described by a single power law with a rather canonical spectral index of $-0.58 \pm 0.02$. Below $1.25$ GHz, the spectrum turned over and the two-point spectral index $\alpha^{1250}_{143.5}$ was  optically thick ($0.74 \pm 0.13$). By the end of our observations, the broadband spectrum had clearly evolved: the LOFAR data point was the brightest of the four flux density measurements, and the spectrum did not turn over in the same pronounced way that the previous epochs showed. Indeed, $\alpha^{1250}_{143.5}$ was now flat and optically thin ($-0.22 \pm 0.11$), and above $1.25$ GHz the spectral index had steepened to $-0.79 \pm 0.03$. A simple explanation for this overall spectral variation could be as postulated in other studies (Section~\ref{introduction}): as the synchrotron-emitting plasmons from the outburst expanded as they travelled away from the Cygnus X-3 system, they became progressively less optically thick at lower frequencies, and subsequently the spectral turnover frequency shifted to lower values as the flare simultaneously decayed at the higher frequencies. However, as we will discuss further in Section~\ref{predictions}, then we would have expected the flare to be brightest at our highest observing frequency of $15$ GHz \citep[e.g.][]{vanderlaan66}.    

At GHz frequencies, both Table~\ref{table:alphas} and Figure~\ref{fig:seds} show that, in general, single-power-law spectral fits are appropriate for most of the epochs. Discrepancies are most apparent for Runs 4 and 6, however: there is either strong or very strong statistical evidence that our model in Equation~\ref{eqnfit} does not appropriately describe the data for these two cases ($P$-values of $3 \times 10^{-3}$ and $3 \times 10^{-6}$ that $\chi^2_{\rm red}$ would be larger for Runs 4 and 6, respectively). For Run 4, it is interesting to note that there was a one-day dip in the $1.25$-GHz flux density a few days before the flare peaked at this frequency. For Run 6, the flare was peaking at $1.25$ GHz, yet had begun to decay at $2.3$ and $15$ GHz. Therefore, the standard delay in the evolution of the flare with decreasing frequency would very likely explain any deviation from a single power law at the epoch of Run 6. This delay would also explain why $\alpha^{2300}_{1250}$, $\alpha^{15000}_{2300}$ and the $1.25$--$15$ GHz fitted spectral index either clearly follow, or show hints of, a general trend of first flattening as the flare peaked at higher frequencies, and then steepening thereafter. The broadband GHz-frequency spectrum was also still steepening at the end of our observing campaign, with the spectral index having not reached a possible terminal value, as seen in some previous studies (e.g. discussion in \citealt[][]{millerjones04}, and references therein). On the contrary, three days after our LOFAR observations ended, the spectrum between $1.25$ and $15$ GHz had evolved significantly, with evidence of turnover owing to rapid decay at $1.25$ GHz (Figure~\ref{fig:light curve}). Further analysis of the GHz-frequency data will be presented in future papers. 

An approximately constant value of $\alpha^{1250}_{143.5} \approx 0.8$ was observed from Runs 1--6. The subsequent evolution of this two-point spectral index between Runs 1--6 and 8, $\Delta\alpha^{1250}_{143.5} \approx -1$, was much more pronounced than the moderate net steepening that occurred at the higher frequencies over the course of the observing campaign ($\Delta\alpha_{\rm fitted} \approx -0.2$). It is interesting to note that while there was a significant steepening in $\alpha^{1250}_{143.5}$ for Run 6 due to the flare peaking at $1.25$ GHz and beginning to decay at $2.3$ GHz, $\alpha^{1250}_{143.5}$ remained at a similar value as determined for the previous runs; we may have expected $\alpha^{1250}_{143.5}$ to become more inverted at this epoch due to the further delay in the flare peaking below $1.25$ GHz. However, it is likely that the comparison of the $\alpha^{1250}_{143.5}$ values for Runs 1--6 is affected by the relative lack of precision in the LOFAR flux density measurements.      

\subsection{Modelling the spectral turnover}\label{light curve comparison3}

\begin{figure*}
\begin{minipage}{0.5\textwidth}
\includegraphics[width=8.4cm]{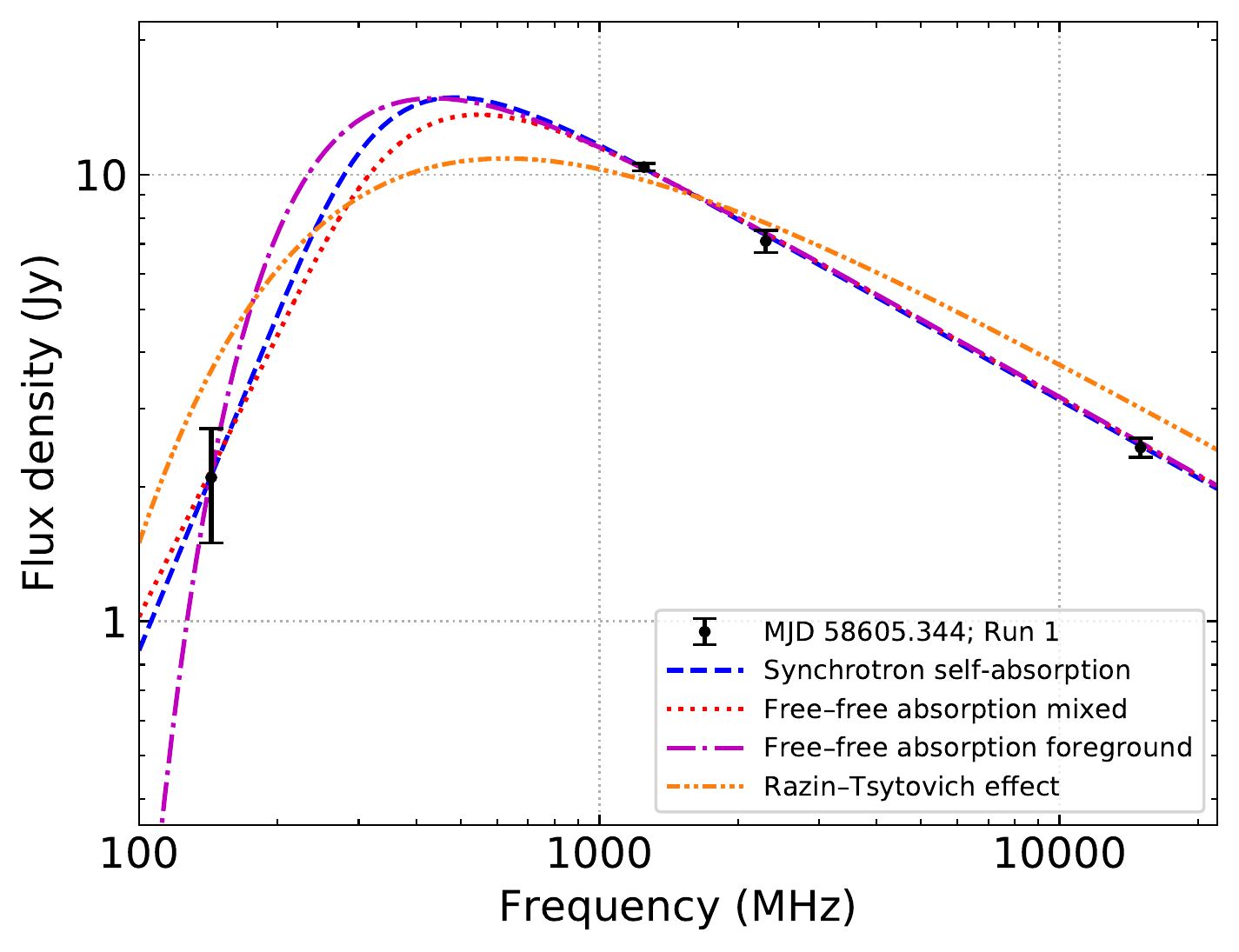}
\end{minipage}%
\begin{minipage}{0.5\textwidth}
\includegraphics[width=8.4cm]{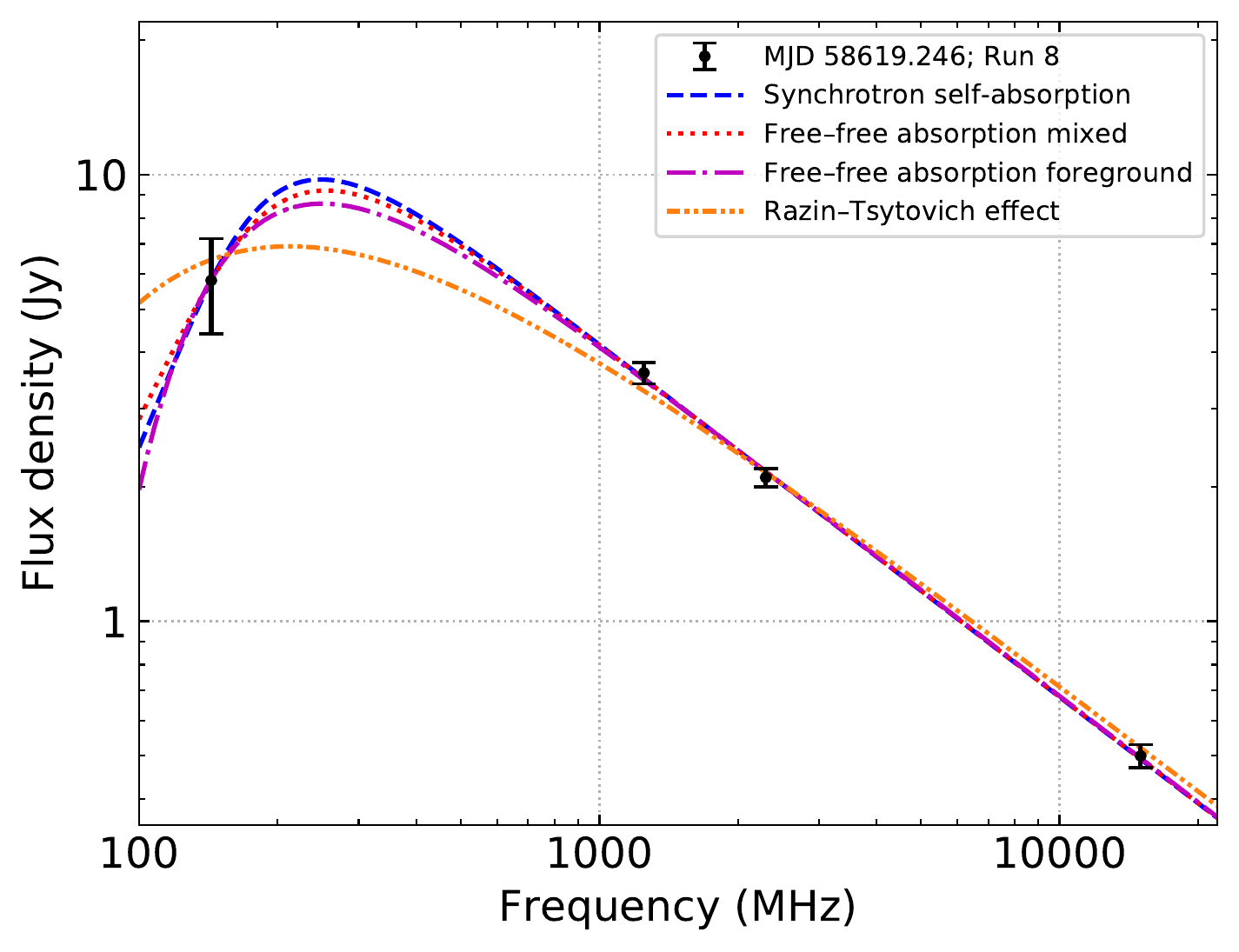}
\end{minipage}
\caption{Broadband radio spectra from Runs 1 and 8 (left and right panels, respectively), overlaid with the model fits described in Equations~\ref{eqn:SSA}--\ref{eqn:RT}. The fitted parameters are given in Table~\ref{table:fits_absorption}. Aside from the Razin--Tsytovich effect for the first epoch (shown in the left panel as the dot-dot-dashed line, which has $\chi^2_{\rm red} = 21$) all models are plausible for both runs. This result emphasizes the need for observational constraints between $143.5$ and $1250$ MHz, in particular below $\sim$$800$ MHz, to better constrain the frequency and flux density of the spectral turnover, and in turn its likeliest cause. See Section~\ref{predictions} for further details.}
\label{fig:seds2}
\end{figure*}

Another point of interest from Figure~\ref{fig:seds} and Table~\ref{table:alphas} is that, in the absence of the LOFAR data, the flare would have been interpreted as being optically thin only. With the LOFAR data included, we could investigate possible mechanisms for spectral turnover. For example, \citet[][]{gregory74} considered four explanations for spectral turnover as part of their modelling study of the 1972 outburst from Cygnus X-3: (i) synchrotron self-absorption, (ii) free--free absorption from thermal plasma mixed with the synchrotron-emitting plasma, (iii) free--free absorption from a foreground screen, and (iv) the Razin--Tsytovich effect (also see e.g. \citealt[][]{millerjones04} and \citealt[][]{koljonen18} for similar investigations of other outbursts). The relevant models are then as follows:
\begin{equation}\label{eqn:SSA}
S_{\nu} = \frac{S_{\tau}}{1-\exp(-1)}\left(\frac{\nu}{\nu_{\tau}}\right)^{2.5}\left(1-\exp\left[-\left(\frac{\nu}{\nu_{\tau}}\right)^{\alpha_{\rm fitted} - 2.5}\right]\right)
\end{equation}
for mechanism (i),
\begin{equation}\label{eqn:FF_mixed}
S_{\nu} = \frac{S_{\tau}}{1-\exp(-1)}\left(\frac{\nu}{\nu_{\tau}}\right)^{2.1}\left(1-\exp\left[-\left(\frac{\nu}{\nu_{\tau}}\right)^{\alpha_{\rm fitted} - 2.1}\right]\right)
\end{equation}
for mechanism (ii),
\begin{equation}\label{eqn:FF_screen}
S_{\nu} = S_{\tau}\left(\frac{\nu}{\nu_{\tau}}\right)^{\alpha_{\rm fitted}}\frac{\exp\left(-[\nu/\nu_{\tau}]^{-2.1}\right)}{\exp(-1)}
\end{equation}
for mechanism (iii), and
\begin{equation}\label{eqn:RT}
S_{\nu} = S_{\rm RT}\left(\frac{\nu}{\nu_{\rm RT}}\right)^{\alpha_{\rm fitted}}\frac{\exp(-\nu_{\rm RT}/\nu)}{\exp(-1)}
\end{equation}
for mechanism (iv). In Equations~\ref{eqn:SSA}--\ref{eqn:FF_screen}, the optical depth $\tau$ is unity at frequency $\nu_{\tau}$, where the flux density is $S_{\tau}$. In Equation~\ref{eqn:RT}, $\nu_{\rm RT}$ is the cutoff frequency for the Razin--Tsytovich effect, below which the flux density is significantly suppressed; $S_{\rm RT}$ is the flux density at this frequency. The values of $\alpha_{\rm fitted}$ are given in Table~\ref{table:alphas}. The frequencies $\nu_{\tau}$ and $\nu_{\rm RT}$ are not where the model fits peak; we denote the peak frequency as $\nu_{\rm p}$. Our observations did rule out a fifth possibility: the spectral slope between $143.5$ and $1250$ MHz ($\alpha^{1250}_{143.5}$ $\approx$ $0.8$ in Runs 1--6) was significantly more inverted than the expected spectral slope ($\alpha = 0.3$) of a low-frequency cutoff in the electron energy spectrum.

We fitted the models shown in Equations~\ref{eqn:SSA}--\ref{eqn:RT} to the data presented in Table~\ref{table:alphas}. Generally speaking, synchrotron self-absorption and free--free absorption (both scenarios for the latter) were plausible absorption mechanisms for all runs apart from Runs 4 and 6, where, as described above, the optically-thin GHz-frequency spectrum was not well fitted by a single power law. The Razin--Tsytovich fit results were generally poor, aside from Run 8. In Figure~\ref{fig:seds2} we show the model fits for Runs 1 and 8, and in Table~\ref{table:fits_absorption} we give the fitted parameter values for these two runs. Table~\ref{table:fits_absorption_appendix} in Appendix~\ref{appendix3} contains the fitting results for all eight runs. Similar to the fitting described in Section~\ref{light curve comparison2}, the $\chi^2_{\rm red}$ values in Tables~\ref{table:fits_absorption} and \ref{table:fits_absorption_appendix} suggest that the flux densities have overestimated uncertainties. However, given that there were only two degrees of freedom for each model fitted, the formal standard deviation for the $\chi^2_{\rm red}$ distribution is again relatively large, and equal to unity. Furthermore, we caveat that our results would be strongly affected by blending of individual flares caused by the prolonged activity, particularly at LOFAR frequencies (additional discussion in Section~\ref{predictions}). 

Consistent with our previous discussion (Section~\ref{light curve comparison2}), regardless of mechanism, $\nu_{\rm p}$ shifted to a lower value over the course of our observations (Tables~\ref{table:fits_absorption} and \ref{table:fits_absorption_appendix}). Filling in the frequency gap between the LOFAR and RATAN-600 data, for example at $325$ and $610$ MHz with the upgraded GMRT (uGMRT), would have provided better constraints on the turnover frequency at each epoch, potentially allowing us to identify the most probable absorption mechanism(s). In the following section, we continue the discussion of this topic, but this time examining the peak flux density of the flare as a function of frequency. 

\begin{table}
 \centering
  \caption{Turnover frequencies, fitted model parameters, and goodness-of-fit statistics for the models described in Section~\ref{light curve comparison3}. A subset of the fits (for Runs 1 and 8) is plotted in Figure~\ref{fig:seds2}. Results for all of the runs can be found in Table~\ref{table:fits_absorption_appendix} in Appendix~\ref{appendix3}. The fit types given below are abbreviated descriptions of mechanisms (i)--(iv) described in Equations~\ref{eqn:SSA}--\ref{eqn:RT}.}
  \begin{tabular}{lcccc}
    \hline
\multicolumn{1}{c}{Run and} & \multicolumn{1}{c}{$\nu_{\rm p}$} & \multicolumn{1}{c}{$\nu_{\tau}$} & \multicolumn{1}{c}{$S_{\tau}$} & \multicolumn{1}{c}{$\chi^2_{\rm red}$} \\ 
\multicolumn{1}{c}{type of fit} & & \multicolumn{1}{c}{or $\nu_{\rm RT}$} & \multicolumn{1}{c}{or $S_{\rm RT}$} & \\ 
& \multicolumn{1}{c}{(MHz)} & \multicolumn{1}{c}{(MHz)} & \multicolumn{1}{c}{(Jy)} & \\
  \hline
1; SSA & $490$ & $360 \pm 30$  & $13.6 \pm 0.7$ & $0.23$  \\
1; FFA mixed & $550$ & $410 \pm 40$  & $12.7 \pm 0.7$ & $0.47$ \\
1; FFA foreground & $440$ & $240 \pm 10$  & $10.3 \pm 0.3$ & $0.53$ \\
1; RT effect & $630$ & $360 \pm 20$  & $9.8 \pm 0.3$ & $21$ \\
\\
8; SSA & $250$ & $200 \pm 20$  & $9.3 \pm 0.7$ & $0.37$  \\
8; FFA mixed & $260$ & $210 \pm 20$  & $8.9 \pm 0.7$ & $0.38$ \\
8; FFA foreground & $250$ & $160 \pm 20$  & $6.7 \pm 0.5$ & $0.44$ \\
8; RT effect & $210$ & $170 \pm 30$  & $6.8 \pm 0.9$ & $1.8$ \\
\hline
\end{tabular}
\label{table:fits_absorption}
\end{table}

\subsection{Comparing the 143.5-MHz flare peak with model predictions}\label{predictions}

\begin{figure}
\epsfig{file=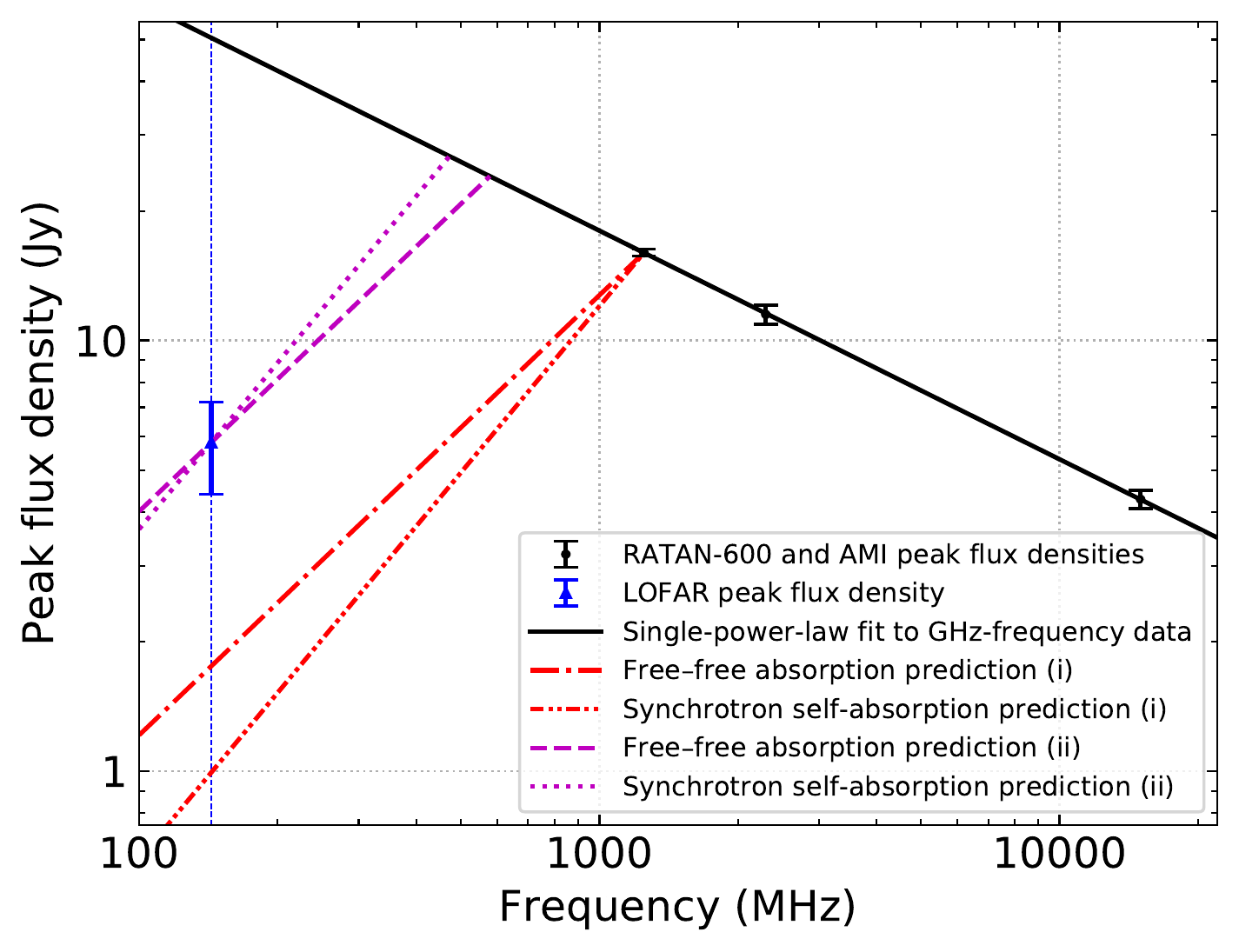,width=8.4cm}
\caption{Peak flux densities of the flaring activity in 2019 May as a function of frequency. The LOFAR peak flux density (Run 8 in Table~\ref{table:alphas}) is shown with a triangle to indicate that it is effectively a lower limit given our limited sampling. The black line is the (extrapolated) single-power-law fit to the GHz-frequency flux densities (Section~\ref{predictions}). As described in Section~\ref{predictions}, we also show predictions from the \citet[][]{marti92} model for both the free--free absorption and synchrotron self-absorption cases: (i) the spectrum turns over immediately below $1.25$ GHz, and (ii) the spectrum turns over at lower frequencies such that the predictions match our measured LOFAR peak flux density. The vertical blue line has been plotted to help indicate the predicted $143.5$-MHz flux densities for (i). In practice, the spectral turnover would be smoother than indicated in this figure.}
\label{fig:seds3}
\end{figure}

We also investigated whether the $143.5$-MHz flare was brighter or fainter than expected given the higher-frequency properties (Figure~\ref{fig:seds3}). First, we made the assumption that the expected quiescent baseline at each frequency (see Sections~\ref{introduction} and \ref{light curve}) was negligible in comparison to the much brighter peak flux density. Similarly to the analysis in Section~\ref{light curve comparison2}, we then fitted a single power law to the radio spectrum of GHz-frequency peak flux densities from Figure~\ref{fig:light curve} ($16.0 \pm 0.3$, $11.50 \pm 0.58$ and $4.28 \pm 0.21$ Jy at $1.25$, $2.3$ and $15$ GHz, respectively), which yielded a spectral index of $-0.53 \pm 0.02$ ($\chi^2_{\rm red}=0.015$). Therefore, this spectrum was also optically thin to at least as low as $1.25$ GHz.     

\citet*[][]{marti92} developed a model for describing outbursts from Cygnus~X-3, featuring twin jets that (i) are laterally expanding, and (ii) into which relativistic electrons are being injected. They found that it could describe the properties of the 1972 September outburst over the frequency range $0.4$--$90$ GHz (i.e. slightly above our LOFAR observing frequency at the lower end of this range). This was in contrast to invoking the often-used \citet[][]{vanderlaan66} model. The peak flux densities of Cygnus X-3 from the 2019 May flare were also not consistent with the \citet[][]{vanderlaan66} synchrotron bubble model, given that (i) the peak flux density decreased with increasing frequency above $1$ GHz, and (ii) the LOFAR peak flux density was significantly brighter than would have been predicted \citep[see e.g. similar results from LOFAR monitoring of the X-ray binary SS 433;][]{broderick18}. Also note that, for example, the continued production of relativistic electrons was hypothesized to explain the increase in flux density with decreasing frequency in the 1991 January 31 outburst from the black-hole X-ray binary GRS 1124$-$68 / Nova Muscae 1991 \citep[][]{ball95}. 

In the \citet[][]{marti92} model, for frequencies where the emission is initially optically thin, 
\begin{equation}
S_{\nu,\,\rm max} \propto \nu^{\alpha_{\rm thin}},
\end{equation}
where $S_{\nu,\,\rm max}$ is the peak flux density of the flare at frequency $\nu$, and $\alpha_{\rm thin}$ is the single-power-law spectral index in the optically thin regime. For frequencies that are initially optically thick,   
\begin{equation}\label{eqn:marti1}
S_{\nu,\,\rm max} \propto \nu^{(6.24 - 5.56\alpha_{\rm thin})/9}
\end{equation}
for the case of free--free absorption (where the thermal gas is mixed with the synchrotron-emitting plasma), and  
\begin{equation}\label{eqn:marti2}
S_{\nu,\,\rm max} \propto \nu^{(15 - 26\alpha_{\rm thin})/(15-14\alpha_{\rm thin})}
\end{equation}
for the case of synchrotron self-absorption. Making the assumption that the 2019 May outburst was well described by this model with $\alpha_{\rm thin}$ = $-0.53$, as calculated above (this value is very similar to $\alpha_{\rm thin} = -0.55$ determined by \citealt[][]{marti92} for the 1972 outburst), we then determined from Equations~\ref{eqn:marti1} and \ref{eqn:marti2} that $S_{\nu,\,\rm max} \propto \nu^{1.02}$ for the case of free--free absorption, and $S_{\nu,\,\rm max} \propto \nu^{1.28}$ for synchrotron self-absorption. If the spectrum was optically thick immediately below $1.25$ GHz, then the $143.5$-MHz flux densities would have been expected to peak at about $1.8$ and $1.0$ Jy for the free--free and synchrotron absorption cases, respectively. These values are both significantly below our measured flux densities in Runs 7 and 8, and the discrepancy would be even larger if our observations did not catch the peak of the flare at $143.5$ MHz. Indeed, if we assume that $S_{143.5,\,\rm max} \gtrsim 5.8$ Jy, then $\alpha^{1250}_{143.5} \lesssim 0.5$, significantly less inverted than would be expected from Equations~\ref{eqn:marti1} and \ref{eqn:marti2}. 

There are at least two possible scenarios to explain the discrepancy. Firstly, the turnover frequency in the radio spectrum of peak flux densities may have been significantly below $1.25$ GHz. Using Equations~\ref{eqn:marti1} and \ref{eqn:marti2}, to ensure consistency with our brightest $143.5$-MHz flux density of $5.8$ Jy, the radio spectrum would have needed to transition from the optically thin to optically thick regime at frequencies of roughly $580$ and $470$ MHz for the cases of free--free absorption and synchrotron self-absorption, respectively. The mid-frequency peak flux densities would have then potentially reached values $> 20$ Jy, not seen before in previous flaring events. For comparison, the maximum $614$-MHz flux density in the \citet[][]{pal09} study of the 2006 outburst was about $7.5$ Jy. Additionally, \citet[][]{anderson72} measured a maximum $408$-MHz flux density of over $7$ Jy in 1972 September; also see the $408$-MHz monitoring results from \citet[][]{bonsignori89}, which covered the period 1983 November -- 1985 September, and in which the highest flux density was $2.2$ Jy. In practice, the transition from the optically thin to optically thick regime would have been smoother than indicated in Figure~\ref{fig:seds3}, both reducing the highest peak flux density at mid frequencies and shifting the transition frequency to a higher value. Nonetheless, the data in Figure~\ref{fig:seds3} (as well as the fitting results in Tables~\ref{table:fits_absorption} and \ref{table:fits_absorption_appendix}) strongly suggest that this flare was very bright at mid frequencies. 

Secondly, the underlying physics may require a different modelling approach, particularly as the overall flaring activity was extended over a period of more than one month. In fact, this more than likely complicates the analysis, as the emission from an ensemble of discrete flares would have become blended together, particularly at low frequencies where the rise times are, generally speaking, significantly longer. If we assume that the effective baseline level at $143.5$ MHz was roughly $2$ Jy before the final bright flare (i.e. this was the remaining low-frequency response from the flaring prior to our observations taking place), then the effective $143.5$-MHz peak flux density was $\gtrsim$ $3.8$ Jy. This could help to reconcile the LOFAR peak flux density with the predictions of the \citet[][]{marti92} model, although a thorough comparison would also involve determining appropriate baseline levels from the previous flaring activity at the higher frequencies too. As an estimate, not including baseline corrections at higher frequencies, the transition from the optically thin to optically thick regime would be shifted upwards to frequencies of roughly $760$ and $600$ MHz for the cases of free--free absorption and synchrotron self-absorption, respectively. Furthermore, although beyond the scope of this paper, an alternative approach could be to use a model such as the one presented in \citet[][]{lindfors07}, which assumes that an outburst can be explained by internal shocks in the jets, and decomposes the activity into a separate number of flaring events. 

We can also see in Figure~\ref{fig:light curve} that the flare peaked about one day later at $1.25$ GHz (MJD 58610) than it did at $2.3$ and $15$ GHz (MJD 58609). The daily, or sometimes longer sampling cadence at all three frequencies meant that it was not possible to further quantify the delay as a function of frequency at and above $1.25$ GHz. It was also difficult to assess with confidence the delay between $2.3$/$15$ GHz and $143.5$ MHz. However, for example, if the $143.5$-MHz flare had peaked at or near the time of our final LOFAR observation, then the delay would have been approximately ten days. This is not unreasonable, given both the results of previous low-frequency studies \citep[][]{bash73,millerjones04,millerjones07,pal09}, and the model delays determined by \citet[][]{marti92}. A very conservative lower limit would be about four days, based on the time gap between the $2.3$/$15$-GHz peak and our Run 7 (but see discussion in Section~\ref{light curve comparison}). 

\subsection{Flare energetics}\label{physics}

We used the analytical framework presented in \citet[][]{fender19}, placing estimates on the minimum energy, power, and magnetic field required for the bright May flare. Their analysis is based on the assumption that the evolution of a radio flare is due to changes in the synchrotron self-absorption optical depth. We estimated that the minimum energy was $E_{\rm min} \sim 10^{44}$ erg, corresponding to a minimum power $P_{\rm min} \sim 10^{38}$ erg s$^{-1}$ and a magnetic field at $E_{\rm min}$ of $B_{\rm min} \sim 40$ mG. We also estimated that the expansion velocity of the flare was $\lesssim 0.6c$, with a size $\lesssim 6 \times 10^{13}$ m and brightness temperature $\gtrsim 4 \times 10^{10}$ K.  Due to our low observing frequency, our minimum energy and power estimates are about an order of magnitude higher than values calculated for Cygnus~X-3 by \citet[][]{fender19}, who reported on $2.3$-GHz and $8.3$-GHz monitoring of flaring in 1994.\footnote{See \citet{waltman95} and \citet{fender97} for full details of that outburst.} As a result, our determined $B_{\rm min}$ is also lower than that calculated by \citet[][]{fender19}. 

Using the rise time and maximum brightness of our LOFAR monitoring as an alternative estimate for the jet energetics \citep[e.g. see][]{fender06}, we obtained similar values for the above quantities,  albeit a factor of a few lower. These values are comparable to estimates using the same method for radio flares from a number of other X-ray binaries \citep[e.g.,][]{fender99,fender06,brocksopp07,curran14,russell19}.

\subsection{A comment on our observing strategy}
 
It could perhaps be regarded as unfortunate that our sampling was at a higher cadence when the $143.5$-MHz flux density was approximately constant at the $\sim$$2$-Jy level. Initially, when we triggered our LOFAR observing campaign, it was not entirely clear from evidence available at the time if the source would continue to flare. Hence, there was a strong consideration to obtain several monitoring observations as quickly as possible. Also, because of the challenges in calibrating and imaging our data (Sections~\ref{observations} and \ref{flux scale accuracy}), there was enough of a latency between observations and initial flux density measurements (over one week) such that we could not easily adjust the observing strategy on the fly. The lessons learned in this study will be valuable in optimising the strategy for future monitoring observations.
 
\section{Conclusions and future work}\label{conclusions}

In this study, we presented and analysed LOFAR high-band observations of the 2019 April--May outburst from the X-ray binary Cygnus~X-3. Moreover, we compared our LOFAR data with contemporaneous observations taken with the RATAN-600 and AMI telescopes at frequencies of $1.25$, $2.3$ and $15$ GHz. Our conclusions are as follows.
\begin{enumerate}
\item Over a two-week observing period (May 2--16), we detected statistically significant variability from Cygnus~X-3. In all eight observations, the source was bright and detected well above the usual quiescent flux density level, with the $143.5$-MHz flux density in each observing run ranging from about $1.7$ to $5.8$ Jy. 
\item An approximately constant initial $143.5$-MHz flux density level of $\sim$$2$ Jy from May 2--7 was suggested to be the delayed and potentially blended low-frequency emission from at least some of the flaring activity that was detected at GHz frequencies prior to our observing campaign taking place.  
\item The subsequent increase in the $143.5$-MHz flux density by nearly a factor of three, to a measured peak of $5.8$ Jy on May 16, was interpreted as the low-frequency equivalent of the bright flare seen on May 6/7 at GHz frequencies. There is a tentative suggestion that the peak $143.5$-MHz flux density was significantly brighter than would be expected given the properties of the flare at GHz frequencies. While there is evidence that the low-frequency peak was delayed by more than four days compared to the peaks at $2.3$ and $15$ GHz on May 6, our light curve sampling was not sufficiently fine enough to determine with certainty when the $143.5$-MHz peak occurred. It is possible that the $143.5$-MHz light curve continued to rise after our observations ended, or the flare peaked between Runs 7 and 8 (i.e. between May 10 and 16).
\item As in other studies of Cygnus~X-3 in outburst, we found that there was a clear evolution in the broadband radio spectrum. In particular, the amount of spectral curvature below $1.25$ GHz significantly decreased ($\Delta\alpha^{1250}_{143.5} \approx -1$), implying that the turnover frequency had shifted to lower frequencies over the course of our LOFAR observing campaign. A simple interpretation is that the outburst became progressively less optically thick at lower frequencies as it progressed. However, we did not have enough radio spectral coverage to conclusively identify whether synchrotron self-absorption or free--free absorption was responsible for the spectral turnover, nor could we determine a fully satisfactory explanation for how the peak flux density varied as a function of frequency. Further modelling and analysis is needed. 
\item We estimated a number of physical properties of the bright flare for which we had the LOFAR coverage, in particular a minimum energy, magnetic field and mean power of roughly $10^{44}$ erg, $40$ mG and $10^{38}$ erg~s$^{-1}$, respectively. Our values are broadly consistent with previous outbursts of both Cygnus X-3 and other X-ray binaries.  
\end{enumerate}

As discussed in Section~\ref{observations}, successful direction-dependent calibration, as well as the subsequent subtraction of Cygnus A from the visibilities, would allow higher-resolution, higher-dynamic-range maps to be constructed. These refined images could then be used to investigate in further detail how much Cygnus~X-3 varied from run to run over the course of our observing campaign. The in-band spectral properties could potentially be better analysed too.   

In the case of the 2019 April--May flaring from Cygnus~X-3, additional activity was detected shortly afterwards in 2019 June \citep[][]{tsubono19b,piano19c}, including an even brighter radio outburst (S. Trushkin, priv. comm.). Further outbursts were detected in 2020 February and June \citep[][]{piano20,trushkin20a,trushkin20b,tsubono20a,tsubono20b,egron20,green20}. For the next giant outburst from Cygnus~X-3, it would be valuable to have high-cadence, low-frequency monitoring over the full duration of the flaring. If Cygnus~X-3 were to reach Jy-level flux densities again in the LOFAR high band during such an outburst, then reducing each monitoring scan in length by a factor of two would still give sufficient $(u,v)$ coverage for imaging \citep[e.g. the strategy used for LOFAR monitoring of the microquasar SS 433 by][]{broderick18}. Additional observations in the LOFAR low band ($30$--$80$ MHz) could also be of interest, particularly to better constrain spectral turnover, and to search for a cutoff frequency in the synchrotron spectrum.      

\section*{Acknowledgements}

We are saddened to report that Dr G. G. Pooley passed away before this paper was published. For many decades Dr Pooley made considerable contributions to a variety of projects, including the monitoring of Cygnus X-3 and many other variable radio sources using the Cambridge $5$-km and Ryle telescopes, and most recently the AMI telescope.

We thank the referee for their helpful comments and suggestions, which improved the presentation of this paper.

This paper is based on data obtained with the International LOFAR Telescope (ILT) under project code LC11\_021. LOFAR \citep[][]{vanhaarlem13} is the Low Frequency Array designed and constructed by ASTRON. It has observing, data processing, and data storage facilities in several countries, that are owned by various parties (each with their own funding sources), and that are collectively operated by the ILT foundation under a joint scientific policy. The ILT resources have benefitted from the following recent major funding sources: CNRS-INSU, Observatoire de Paris and Universit\'{e} d'Orl\'{e}ans, France; BMBF, MIWF-NRW, MPG, Germany; Science Foundation Ireland (SFI), Department of Business, Enterprise and Innovation (DBEI), Ireland; NWO, The Netherlands; The Science and Technology Facilities Council, UK; Ministry of Science and Higher Education, Poland.  

Observations with the RATAN-600 radio telescope are supported by the Ministry of Science and Higher Education of the Russian Federation. For AMI, we acknowledge support from the European Research Council under grant ERC-2012-StG-307215 LODESTONE. We thank the staff of the Mullard Radio Astronomy Observatory for support in the operation of AMI.

TDR acknowledges support from a Netherlands Organisation for Scientific Research (NWO) Veni Fellowship and a financial contribution from the agreement ASI-INAF n.2017-14-H.0. DRAW and JSB were supported by the Oxford Centre for Astrophysical Surveys, which is funded through generous support from the Hintze Family Charitable Foundation. 

We thank the ASTRON Radio Observatory, particularly Sander ter Veen and Pietro Zucca, for promptly setting up and scheduling the observations described in this paper, preprocessing the data, and providing advice on some data reduction issues. We also thank the ASTRON Radio Observatory for kindly granting us processing time on the CEP3 cluster,  Andr{\'e} Offringa and Sarrvesh Sridhar for helpful feedback on a {\sc wsclean} imaging question, Huib Intema and Anna Kapi{\'n}ska for useful suggestions regarding the accuracy of the low-frequency flux density scale, and James Miller-Jones for providing us with a $330$-MHz WSRT map of the Cygnus~X-3 field. 

This research has made use of the VizieR catalogue access tool, CDS, Strasbourg, France (DOI: 10.26093/cds/vizier). The original description of the VizieR service was published in A\&AS 143, 23 \citep*[][]{ochsenbein00}. This research has made use of the SIMBAD database, operated at CDS, Strasbourg, France \citep[][]{wenger00}. This research has made use of NASA's Astrophysics Data System Bibliographic Services. This project also made use of {\sc kern} \citep[][]{molenaar18}, {\sc kvis} \citep[][]{gooch95}, {\sc matplotlib} \citep[][]{hunter07}, {\sc numpy} \citep[][]{oliphant06}, {\sc scipy} \citep[][]{virtanen20}, {\sc topcat} \citep[][]{taylor05}, and Overleaf (\url{http://www.overleaf.com}).

\section*{Data availability}

The preprocessed LOFAR data sets used in this study are available under project code LC11\_021 in the LOFAR Long-Term Archive (\url{https://lta.lofar.eu}). Calibrated LOFAR data and images will be shared on reasonable request to the corresponding author. The RATAN-600 and AMI flux density data sets will be presented in future papers; please contact the corresponding author for further information. TGSS data products are available at \url{http://tgssadr.strw.leidenuniv.nl/doku.php}.   


\appendix

\section{Correcting the flux density scale of the LOFAR data}\label{appendix1}

\begin{figure*}
\begin{minipage}{1.0\textwidth}
\centering
\includegraphics[height=5.0cm]{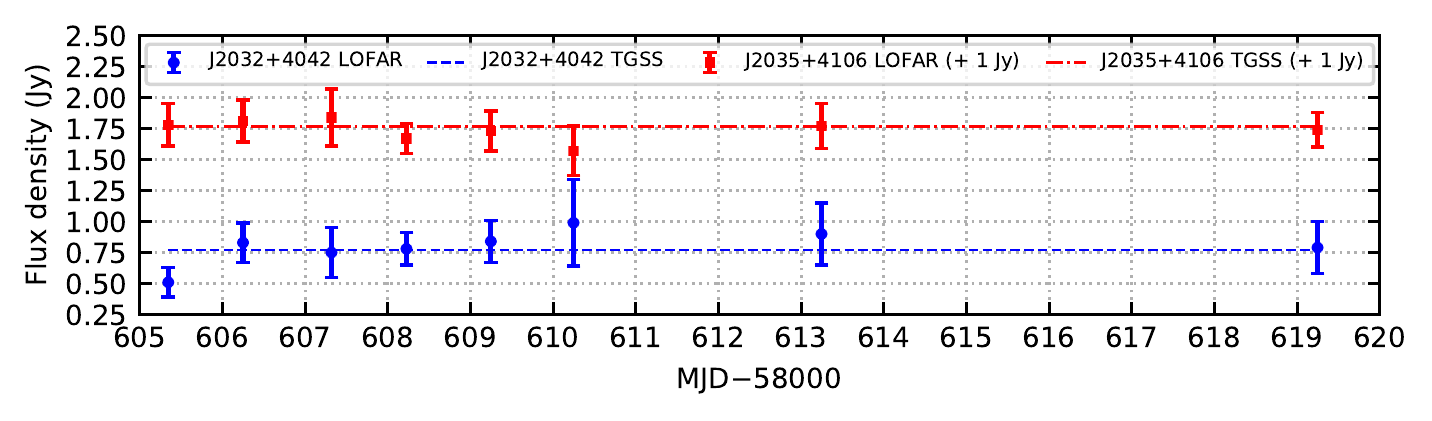}
\end{minipage}%
\caption{The $143.5$-MHz light curves of the two field sources used to investigate the accuracy of our bootstrapping procedure (see Figure~\ref{fig:example}). The dashed lines are the $147.5$-MHz TGSS catalogue values. Flux densities for J203533.2$+$410645 have been shifted upwards by $1$ Jy for the purposes of clarity. Error bars are $\pm1\sigma$. As in Figure~\ref{fig:light curve}, the assumed $20$ per cent uncertainty in the absolute flux density scale is not accounted for in the plotted error bars.}
\label{fig:appendix}
\end{figure*}

To bootstrap the flux density scale from TGSS, we used five well-spaced, bright comparison sources with $S_{147.5} > 1$ Jy that are within $2$\degr\,(i.e. the primary beam half power point) of Cygnus~X-3. In order of decreasing brightness, these sources are J203556.7$+$421747, J203745.3$+$391533, J202753.4$+$423158, J203934.1$+$402539 and J202542.8$+$415615 (marked in Figure~\ref{fig:example} with circles). Source finding in our LOFAR maps was carried out with {\sc pybdsf}; we then calculated the TGSS/LOFAR integrated flux density ratio for each of our comparison sources, per run. For a given run, the mean flux density ratio was used as the bootstrapping correction factor, and the associated uncertainty was the standard deviation of the flux density ratios. For Run 6, J202542.8$+$415615 was removed from the comparison as the signal-to-noise ratio was too low ($< 5$) in the LOFAR map. The correction factors to multiply the LOFAR flux densities by ranged from $1.44$ to $2.33$, with relative $1\sigma$ uncertainties ranging from $9$ to $22$ per cent (Table~\ref{table:obs}). No significant evidence was found to suggest that the correction factors were correlated with either the radial distance from the field centre, or the position within the field. We did not correct for the different central frequencies of our data and TGSS (i.e. $143.5$ and $147.5$ MHz), because this was a second-order effect compared to the scatter in each correction factor. We also did not weight the data when deriving the correction factors: this gave a more conservative estimate of the scatter associated with the bootstrapping. Finally, given the two-week time-scale of our observing campaign, the derived corrections were very unlikely to have been affected by any variability from the comparison sources \citep[e.g. see results from][]{bell19}.

To demonstrate confidence in the bootstrapping procedure, in Figure~\ref{fig:appendix} we show the corrected $143.5$-MHz light curves of two moderately bright TGSS sources that are relatively close to Cygnus~X-3 on the sky (marked in Figure~\ref{fig:example} with ellipses). These sources are J203214.1$+$404223 ($0$\fdg$25$ from Cygnus~X-3; TGSS catalogued flux density $S_{147.5} = 769 \pm 79$ mJy) and J203533.2$+$410645 ($0$\fdg$61$ from Cygnus~X-3; $S_{147.5} = 766 \pm 78$ mJy). As before, {\sc pybdsf} was used for source finding and flux density measurements in the LOFAR maps. After bootstrapping, these two sources had flat light curves (within the uncertainties) during our observing campaign. Moreover, the inverse-variance-weighted mean $143.5$-MHz flux densities, $740 \pm 60$ and $730 \pm 60$ mJy, respectively (where each uncertainty is the standard error of the weighted mean), were consistent with the TGSS catalogue values stated above. Note that there was still an agreement in both cases if the dominant source of uncertainty in the TGSS catalogued flux densities, that is the assumed $10$ per cent uncertainty in the flux density scale \citep[][]{intema17}, was removed. Furthermore, after using the same $\chi^2$ test as that described in Section~\ref{light curve}, we found no evidence for deviation from a constant, flat light curve for the two comparison sources. 

Given the proximity of Cygnus~A to our target, the standard $10$ per cent TGSS calibration uncertainty may not necessarily have been applicable in this study (H. Intema, priv. comm.). We were therefore more conservative and assumed the TGSS flux density scale accuracy to be $20$ per cent for our target field. As an additional reliability check, we took the five bright comparison sources that we used to bootstrap the flux density scale from TGSS, and found their higher-frequency counterparts in the $1400$-MHz NRAO VLA Sky Survey \citep[NVSS;][]{condon98}. The median TGSS--NVSS two-point spectral index ($\alpha^{1400}_{147.5}$) of these sources, $-0.79$, is encouragingly close to the median for the global TGSS--NVSS cross-correlation \citep[$-0.73$;][]{intema17}.

\section{Investigating the possible contribution to the LOFAR flux densities from nearby extended emission}\label{appendix2}

\citet[][]{sanchezsutil08} reported on the detection of low-surface-brightness, arcminute-scale emission in the vicinity of Cygnus~X-3. The integrated flux density at $5$ GHz is $133$ mJy, and $\alpha \approx -0.5$. This emission would have been significantly blended with Cygnus~X-3 at the angular resolution of our LOFAR observations. Extrapolating the $5$-GHz flux density to $143.5$ MHz using $\alpha = -0.5$ gives a $143.5$-MHz flux density of roughly $800$ mJy. However, it is unclear whether this emission follows a single-power-law spectrum to LOFAR frequencies. On the one hand, no corresponding detection was seen in TGSS ($3\sigma$ upper limit $\approx$ $30$ mJy beam$^{-1}$; angular resolution $25$ arcsec), but this could have possibly been due to limitations in the low-surface-brightness sensitivity of this survey \citep[][]{intema17}. On the other hand, \citet[][]{millerjones07} detected diffuse emission in the vicinity of Cygnus~X-3 at a frequency of $330$ MHz (left panel of their figure 3; synthesized beam $93 \times 59$ arcsec$^{2}$ and PA $3$\fdg$3$); also see the $325$-MHz detection in \citet[][]{benaglia21}. We inspected the corresponding image file (J. Miller-Jones, priv. comm.), finding that the surface brightness increased at the same coordinates as the feature reported by \citet[][]{sanchezsutil08}, with a brightness level of about $15$--$25$ mJy beam$^{-1}$. A very crude, non-background-corrected estimate of an upper limit for the $330$-MHz integrated flux density of the extended emission is $\sim$$90$ mJy. This upper limit is well below what would be expected based on a single-power-law extrapolation from $5$ GHz assuming $\alpha = -0.5$, suggesting that the spectrum of the extended emission has turned over at low frequencies due to one or more absorption processes. The corresponding flux density at $143.5$ MHz could therefore be even lower, well within the uncertainties for the Cygnus X-3 measurements reported in Table~\ref{table:obs}. 

\section{Modelling the spectral turnover: full set of results}\label{appendix3}

In Table~\ref{table:fits_absorption_appendix}, we present the results from the spectral turnover modelling that we described in Section~\ref{light curve comparison3}, for all eight observing runs. 

\begin{table}
 \centering
  \caption{Spectral modelling results for all eight runs. See the caption in Table~\ref{table:fits_absorption} for further details.}
  \begin{tabular}{lcccc}
    \hline
\multicolumn{1}{c}{Run and} & \multicolumn{1}{c}{$\nu_{\rm p}$} & \multicolumn{1}{c}{$\nu_{\tau}$} & \multicolumn{1}{c}{$S_{\tau}$} & \multicolumn{1}{c}{$\chi^2_{\rm red}$} \\ 
\multicolumn{1}{c}{type of fit} & & \multicolumn{1}{c}{or $\nu_{\rm RT}$} & \multicolumn{1}{c}{or $S_{\rm RT}$} & \\ 
& \multicolumn{1}{c}{(MHz)} & \multicolumn{1}{c}{(MHz)} & \multicolumn{1}{c}{(Jy)} & \\
  \hline
1; SSA & $490$ & $360 \pm 30$  & $13.6 \pm 0.7$ & $0.23$  \\
1; FFA mixed & $550$ & $410 \pm 40$  & $12.7 \pm 0.7$ & $0.47$ \\
1; FFA foreground & $440$ & $240 \pm 10$  & $10.3 \pm 0.3$ & $0.53$ \\
1; RT effect & $630$ & $360 \pm 20$  & $9.8 \pm 0.3$ & $21$ \\
\\
2; SSA & $530$ & $400 \pm 30$  & $13.9 \pm 0.6$ & $0.15$  \\
2; FFA mixed & $600$ & $460 \pm 40$  & $13.0 \pm 0.6$ & $0.44$ \\
2; FFA foreground & $450$ & $250 \pm 10$  & $10.9 \pm 0.3$ & $0.39$ \\
2; RT effect & $700$ & $420 \pm 20$  & $10.1 \pm 0.2$ & $28$ \\
\\
3; SSA & $480$ & $340 \pm 40$  & $13.9 \pm 0.7$ & $0.038$  \\
3; FFA mixed & $540$ & $390 \pm 50$  & $13.2 \pm 0.7$ & $0.17$ \\
3; FFA foreground & $450$ & $230 \pm 10$  & $10.2 \pm 0.3$ & $0.22$ \\
3; RT effect & $630$ & $320 \pm 30$  & $10.2 \pm 0.3$ & $16$ \\
\\
4; SSA & $550$ & $360 \pm 30$  & $11.2 \pm 0.4$ & $4.1$  \\
4; FFA mixed & $640$ & $420 \pm 40$  & $10.7 \pm 0.4$ & $3.9$ \\
4; FFA foreground & $500$ & $230 \pm 10$  & $7.9 \pm 0.2$ & $4.0$ \\
4; RT effect & $870$ & $360 \pm 20$  & $8.1 \pm 0.2$ & $21$ \\
\\
5; SSA & $530$ & $370 \pm 30$  & $17.3 \pm 0.7$ & $0.13$  \\
5; FFA mixed & $600$ & $430 \pm 40$  & $16.3 \pm 0.7$ & $0.34$ \\
5; FFA foreground & $470$ & $240 \pm 10$  & $12.8 \pm 0.3$ & $0.33$ \\
5; RT effect & $760$ & $380 \pm 20$  & $12.5 \pm 0.3$ & $21$ \\
\\
6; SSA & $500$ & $390 \pm 40$  & $21.1 \pm 1.2$ & $12$  \\
6; FFA mixed & $550$ & $430 \pm 40$  & $19.7 \pm 1.2$ & $13$ \\
6; FFA foreground & $430$ & $250 \pm 10$  & $16.7 \pm 0.5$ & $13$ \\
6; RT effect & $610$ & $400 \pm 20$  & $15.1 \pm 0.4$ & $42$ \\
\\
7; SSA & $340$ & $270 \pm 20$  & $15.6 \pm 0.9$ & $1.1$  \\
7; FFA mixed & $360$ & $290 \pm 30$  & $14.7 \pm 0.9$ & $1.2$ \\
7; FFA foreground & $340$ & $200 \pm 10$  & $11.3 \pm 0.5$ & $1.4$ \\
7; RT effect & $380$ & $260 \pm 30$  & $10.4 \pm 0.6$ & $8.2$ \\
\\
8; SSA & $250$ & $200 \pm 20$  & $9.3 \pm 0.7$ & $0.37$  \\
8; FFA mixed & $260$ & $210 \pm 20$  & $8.9 \pm 0.7$ & $0.38$ \\
8; FFA foreground & $250$ & $160 \pm 20$  & $6.7 \pm 0.5$ & $0.44$ \\
8; RT effect & $210$ & $170 \pm 30$  & $6.8 \pm 0.9$ & $1.8$ \\
\hline
\end{tabular}
\label{table:fits_absorption_appendix}
\end{table}

\bsp	
\label{lastpage}
\end{document}